\documentclass[preprint,aps]{revtex4}
\usepackage{amsmath, amsthm, amssymb}
\usepackage{graphicx}
\usepackage{bm}
\newcommand{\beq}{\begin{equation}}
\newcommand{\enq}{\end{equation}}
\newcommand{\bea}{\begin{eqnarray}}
\newcommand{\ena}{\end{eqnarray}}
\newcommand{\rr}{{\bf r}}
\newcommand{\aos}{a_{\rm osc}}
\newcommand{\tC}{\tilde{C}}
\newcommand{\hw}{\hbar\omega_{\perp}}
\newcommand{\Ecore}{\omega_{\rm core}}
\newcommand{\vcore}{v_{\rm core}}

\def\vec#1{{\bm #1}}

\begin{document}

\title{Dynamical stability of a doubly quantized vortex in a 
three-dimensional condensate}
\author{Emil Lundh}
\affiliation{Centre of Mathematics for Applications, 
P.O. Box 1053 Blindern, NO-0316 Oslo, Norway}
\affiliation{Department of Physics, Ume{\aa} University, 
SE-90187 Ume{\aa}, Sweden\footnote{Present address.}}
\author{Halvor M.\ Nilsen}
\affiliation{Centre of Mathematics for Applications, 
P.O. Box 1053 Blindern, NO-0316 Oslo, Norway}

\begin{abstract}
The Bogoliubov equations are solved for a three-dimensional 
Bose-Einstein condensate containing a doubly quantized vortex, 
trapped in a harmonic potential. 
Complex frequencies, signifying dynamical instability, are found 
for certain ranges of parameter values. The existence of 
alternating windows of stability and instability, respectively, 
is explained qualitatively and quantitatively using variational 
calculus and direct numerical solution. 
It is seen that the windows of 
stability are much smaller for a cigar shaped condensate than for a 
pancake shaped one, which 
is consistent with 
the findings of recent experiments.
\end{abstract}
\maketitle
\section{Introduction}
One of the hallmarks of superfluid flow is the quantization of fluid 
circulation. Its manifestation through the formation of quantized vortices
is a long studied subject in superconductors, superfluid Helium and 
trapped atomic gases \cite{pethick2001}. Although a vortex can in 
principle carry any number of circulation quanta, it is well known that 
in an infinite homogeneous system, only singly quantized vortices are
energetically allowed. 
In such a geometry, a vortex with quantum number larger than 
unity, frequently called a giant vortex, has higher energy than the 
corresponding number of separated 
vortices with single quanta while carrying the same angular momentum.

For Bose-Einstein condensates trapped in magnetic potentials, the 
situation is more complicated since the size of a vortex core can be 
comparable to the total system size. Still, it is an accidental fact 
that quantum numbers larger than unity are energetically 
unfavorable in Bose-Einstein condensates in parabolic potentials, 
just as in the homogeneous case. 
This result hinges on the fact that in the 
noninteracting limit, the lowest energy levels with a given angular 
momentum are vastly degenerate; interactions lift the degeneracy and 
favor a state with a diluted density profile, which is a lattice of 
singly quantized vortices. The situation is different if the trapping 
potential is steeper than harmonic, in which case giant vortices are 
energetically favorable in the limit of weak interactions \cite{lundh2002}.

The related but separate issue of 
{\it dynamical} stability of doubly quantized vortices has become 
especially relevant lately since the problem can be readily probed 
by experimental means \cite{shin2004}. Theoretically, it 
is known that in the two-dimensional limit such a vortex is 
dynamically unstable towards splitting into two
only in certain windows of parameter 
space, when the interaction strength lies within certain intervals 
\cite{pu1999}. These intervals have been shown to be bounded by zeros 
of eigenvalues for the corresponding static stability problem 
\cite{jackson2005}. However, the situation is entirely different in 
three dimensions: in the strongly cigar shaped limit, instabilities are 
expected to lie densely in phase space, because of the closely spaced 
energy levels in the third direction
\cite{mottonen2003,mottonen2006,mateo2006}.
In this paper, we perform a systematic study of the Bogoliubov 
excitations in 
two- and three-dimensional condensates containing a doubly 
quantized vortex. The paper is organized as follows. In 
Sec.\ \ref{sec:method} we discuss the equations and the structure of 
the stability problem. In Sec.\ \ref{sec:twodim} 
we address the
two-dimensional problem and show how the 
numerically
observed parameter dependence can be understood and approximated 
analytically. 
Section \ref{sec:results} contains an overview of the numerical 
results for the three-dimensional condensate. 
In Sec.\ \ref{sec:cigar} we discuss in greater detail 
the limit of a cigar shaped 
condensate. Section \ref{sec:experiment} relates the present 
results to experimental findings.
We calculate the instability
regions for an anharmonic trap in section \ref{sec:anharmonic}, 
to investigate the stability in a more general case.
Finally, 
in Sec.\ \ref{sec:conclusions} we summarize and conclude.

\section{Stability problem for a condensed Bose gas}
\label{sec:method}

The system under study is a Bose-condensed atomic gas that is trapped 
in a cylindrically symmetric harmonic potential. At zero
temperature in the dilute limit it is described by a condensate 
wavefunction $\Psi(r,\theta,z,t)$ that obeys the Gross-Pitaevskii 
equation 
\beq
\label{gpe}
\left[-\frac{\hbar^2}{2m}\nabla^2+ V(\rr)
+U_0|\Psi|^2\right]\Psi
= \mu\Psi.
\enq
The eigenvalue $\mu$ is the chemical potential, and the 
wavefunction $\Psi$ is normalized to the number 
of atoms $N$. 
The trapping potential is assumed to be of the form
\beq
V(\rr) = \frac{m\omega_{\perp}^2}{2}(r^2+\lambda^2 z^2).
\enq
The anisotropy of the trap is governed by the ratio $\lambda$ 
between the axial and radial trapping strengths, so that when $\lambda$ 
is larger than unity, the atomic cloud resembles a pancake 
and when 
$\lambda<1$, it is cigar shaped. We shall see that the system behaves very 
differently in these two limits. 
The inter-particle interactions are parametrized by an s-wave scattering 
length $a$, so that $U_0=4\pi\hbar^2a/m$. Combining these parameters into 
dimensionless quantities, we see that the physics of the system is 
entirely determined by $\lambda$ and the effective 
coupling strength $C=4\pi Na/\aos$, where $\aos$ is the harmonic 
oscillator length, $\aos^2=\hbar/(m\omega_{\perp})$. 
Since the 
chemical potential $\mu$ is a monotonically increasing function of 
$C$ 
(although explicit expressions can be found only in limiting cases), 
it is possible to switch to the pair of parameters 
($\lambda$, $\mu$) instead of ($\lambda$, $C$). This 
parameter choice is found to be more helpful in explaining the 
physics of the problem and is therefore 
used in most of this paper.
We now switch to dimensionless units in which energy is measured in 
units of $\hw$, frequency in units of $\omega_{\perp}$ and length
in units of $\aos$. The Gross-Pitaevskii equation becomes
\beq
\left[-\frac{1}{2}\nabla^2+ \frac12(r^2+\lambda^2z^2)
+C|\Psi|^2\right]\Psi
= \mu\Psi.
\enq
We shall, however, find it natural 
to reinsert units 
when discussing physical scales throughout the paper.

For a given stationary solution $\Psi_0$ of the Gross-Pitaevskii 
equation with eigenvalue $\mu$, the small-amplitude excitations of the 
condensate are defined through the ansatz
\beq
\label{psi_plus_bog}
\Psi(\rr,t) = \left[\Psi_0(\rr)+\sum_{n}\left(u_n(\rr) e^{-i\omega_n t} 
+ v_n(\rr)^* e^{i \omega_n t}\right)\right]e^{-i\mu t},
\enq
where $u_n$ and $v_n$ are the eigenvectors and $\omega_n$ the 
eigenvalues of the Bogoliubov equations, 
\beq
\label{bogoliuboveqn}
B\varphi_n(\rr) = \omega_n\varphi_n(\rr),
\enq
where $\varphi_n(\rr)=(u_n(\rr),v^*_n(\rr))^T$, and the linear 
operator $B$ is defined by
\bea
\label{bogoliubovmat}
B = \left( \begin{array}{cc}
-\frac{1}{2}\nabla^2+V(\rr)-\mu+2C|\Psi|^2 & C\Psi^2 \\
- C(\Psi^*)^2 &-\left(
-\frac{1}{2}\nabla^2+V(\rr)-\mu + 2C|\Psi|^2 \right)
\end{array}\right).
\ena
$B$ is non-Hermitian and may have complex 
eigenvalues. If this is the case, the system 
is dynamically unstable and the corresponding modes will grow 
exponentially,
as seen from Eq.\ (\ref{psi_plus_bog}). 
For the problem at hand, where $\Psi$ is assumed to 
contain a doubly quantized vortex at the origin, it is known in the 
two-dimensional limit that there exist intervals of the coupling 
constant $C$ where exactly one pair of
eigenvalues is complex; in these intervals the doubly quantized vortex 
is unstable towards splitting, 
and between them it is dynamically stable. On the other hand, in the 
strongly cigar shaped limit, the experiment of Ref.\ \cite{shin2004} saw 
instability for all coupling strengths. In this paper, we shall see 
how these two situations emerge as the extreme limits of the general 
three-dimensional problem.

We write the matrix element of a two-by-two, possibly space-dependent, 
matrix $A$ between two Bogoliubov 
eigenvectors as 
$\langle \varphi_m | A | \varphi_n \rangle = 
\int d^3r \varphi_m(\rr)^{\dagger} A(\rr) \varphi_n(\rr)$.
The ``inner product''
of two Bogoliubov eigenvectors $\varphi_n$ is 
defined with the help of the Pauli matrix 
$\sigma_z={\rm diag}(1,-1)$ as 
\beq
\langle \varphi_m | \sigma_z | \varphi_n \rangle = 
\int d^3r \left(u_m(\rr)^*u_n(\rr)-v_m(\rr)^*v_n(\rr)\right).
\enq 
The ``norm'' of a Bogoliubov eigenvector $\varphi_n$ with respect to 
this ``inner product'', $\langle \varphi_n | \sigma_z |\varphi_n\rangle$, 
can be either positive, negative or zero, since it obeys the relation
\beq
(\omega_n-\omega_n^*)\langle \varphi_n | \sigma_z | \varphi_n \rangle = 0.
\enq
If $\omega_n$ is complex, the norm must be zero. In the case of 
real $\omega_n$, 
the norm 
may be either positive or negative and we shall fix 
normalization such that its absolute value is unity for all 
modes with real eigenvalue. 
Note that for every solution $\varphi_n=(u_n,v^*_n)^T$ of the 
Bogoliubov equation with eigenenergy $\omega_n$, there exists 
a solution $\tilde\varphi_n=(v^*_n,u_n)^T$ with eigenvalue $-\omega_n^*$. 
If $\Psi$ is the ground state of the Gross-Pitaevskii equation 
in the absence of rotation, 
the norm of each mode $\varphi_n$ is equal to the sign of 
the corresponding $\omega_n$, but that result does not hold for a 
general $\Psi$. The existence of modes with positive norm and 
negative energy (and necessarily also vice versa) signifies that 
there exist states with lower energy than $\Psi$
\cite{svidzinsky1998,McPeake2002}.

If the problem is cylindrically symmetric and the condensate has an 
angular momentum $M$ per particle, the excitations can be labeled by 
an angular momentum quantum number 
$m$ relative to that of the condensate, such that 
\bea
u(r,\theta,z) = e^{i (M+m) \theta}u(r,z),\nonumber\\
v(r,\theta,z) = e^{i (M-m) \theta}v(r,z).
\ena
The Bogoliubov eigenvalue problem thus splits up into a block diagonal 
matrix where the blocks corresponding to different $m$ are decoupled. 
This allows us to treat each $m$ value separately; 
the $m$'th sector of the Bogoliubov matrix takes on the form
\begin{equation}
\label{cyl_bog_matrix}  
B =  \left[
  \begin{array}{c c}
    D_{1} +V_d(r,z)  &
  V_c(r,z)\\
 - V_c(r,z) &
-\left(D_{2} 
  +V_d(r,z)\right)\\
  \end{array}
  \right],
\end{equation}
where 
\bea
D_1 &=& \frac{1}{2} \left(-\frac{\partial^2}{\partial r^2}-
  \frac{1}{r}\frac{\partial}{\partial r}
  +\frac{(M+m)^2}{r^2}
+\frac{\partial^2}{\partial z^2}
  \right),
\nonumber\\
D_2 &=& \frac{1}{2} \left(-\frac{\partial^2}{\partial r^2}-
  \frac{1}{r}\frac{\partial}{\partial r} 
  +\frac{(M-m)^2}{r^2}
  +\frac{\partial^2}{\partial z^2}
  \right),
\ena
and the diagonal and off-diagonal potentials are defined as 
\bea
V_d(r,z) &=& V(r,z)-\mu+2C|\Psi(r,z)|^2,
\nonumber\\
V_c(r,z)&=&C\Psi(r,z)^2.
\ena
For the problem at hand, the $m=2$ sector will turn out to be 
especially interesting, as we shall see in the following sections.

Let us now study what happens when two Bogoliubov eigenvalues 
collide as a control parameter is changed. We will see that they 
either become complex, signifying a dynamical instability, or 
undergo an avoided crossing. 
Consider two 
Bogoliubov amplitudes $\varphi_j$, with $j=1,2$, that solve 
the Bogoliubov equations for some pair of parameter values $\lambda$, 
$C$. 
We assume that the corresponding eigenvalues $\omega_j$ are real. Now 
slightly increase $C$ to a value $C'=C + \delta C$, so that the 
ground-state solution $\Psi$ becomes $\Psi'=\Psi+\delta\Psi$, and 
the Bogoliubov matrix is correspondingly written $B'=B+\delta B$. 
Consider the restricted Bogoliubov problem in the truncated basis 
formed by $\varphi_1, \varphi_2$. (The validity of this approach 
shall shortly be determined.)  Writing the ansatz as 
$\varphi = a_1 \varphi_1 + a_2\varphi_2$, 
one obtains the
eigenvalue problem for the vector $\vec{a}=(a_1,a_2)^T$,
\beq
\label{twobytwomatrix}
\left(\begin{array}{ll}
\langle \varphi_1|\sigma_z|\varphi_1\rangle & 0 \\
0 & \langle \varphi_2|\sigma_z|\varphi_2\rangle \end{array}\right)
\left(\begin{array}{ll}
\omega_1+\delta B_{11} & \delta B_{12} \\
{\delta B_{21}} & \omega_2+\delta B_{22} \end{array}\right)
\vec{a} = \omega \vec{a},
\enq
where 
$\delta B_{i j}=\langle\varphi_i|\sigma_z\delta B|\varphi_j\rangle$. 
Since $\sigma_z\delta B$ is a Hermitian matrix, we have 
$\delta B_{21}={\delta B_{12}}^*$. 
The eigenvalues are
\bea
\label{complexmodes}
\omega_{\pm} &=& \frac{\omega_1+\delta B_{11}+\omega_2+\delta B_{22}}{2} 
\nonumber\\
&\pm& 
\left[\left(\frac{\omega_1-\omega_2+\delta B_{11}-\delta B_{22}}{2}\right)^2
+\langle \varphi_1|\sigma_z|\varphi_1\rangle
\langle \varphi_2|\sigma_z|\varphi_2\rangle|\delta B_{12}|^2\right]^{1/2}.
\ena
It is seen that the eigenvalues can be substantially altered only if 
the energy difference $\omega_1-\omega_2$ is 
of the same order of magnitude as 
the perturbations $\delta B_{i j}$; thus one only has to care about 
the coupling between nearby energy levels, so the truncation of the 
basis to two states is justified. 
When the two modes have equal norm, they experience an avoided 
crossing.
Conversely, a condition for the eigenvalues to be complex is 
that the two modes have different norm, 
$\langle \varphi_1|\sigma_z|\varphi_1\rangle
\langle \varphi_2|\sigma_z|\varphi_2\rangle=-1$. 
Thus it can be concluded that a dynamical instability is formed by the 
mixing of a Bogoliubov mode of positive norm with one of negative 
norm 
(states with opposite
Krein signature in the language 
of Hamiltonian systems \cite{Arnold68});
when their energies coincide they go over to a pair of zero-norm 
modes with
eigenvalues
that are complex conjugates of each other. We will see plenty of 
examples of this phenomenon in the following. 

For the numerical solution of the Bogoliubov equations, we employ 
the following method.  
At each point in the ($\lambda$, $\mu$) phase space, the stationary 
solution of the Gross-Pitaevskii equation (\ref{gpe}) is sought under 
the condition that it contains a doubly quantized vortex 
along the $z$ axis,
so that the wave function is assumed to be of the form 
\beq
\Psi(r,\theta,z) = e^{2i\theta} \Psi(r,z).
\enq
After that, the Bogoliubov problem is a matter of finding the 
eigenvalues to a $2n\times 2n$ matrix, where $n$ is the size of the 
numerical grid. The spatial coordinates are discretized in a 
discrete-variable representation (DVR) \cite{McPeake2002,baye1986}
using a
Laguerre mesh for the radial direction and a Hermite mesh for the 
axial direction. For the solution of the two-dimensional problem 
reported in 
Secs.\ \ref{sec:twodim} and \ref{sec:anharmonic}, 
the $z$ dependence drops out and 
a one-dimensional Laguerre mesh is employed.

\section{Two-dimensional case revisited}
\label{sec:twodim}

We first consider the case of a two-dimensional condensate, which can be 
realized in a strongly pancake-shaped 
trap, i.~e.\ in the limit of large 
$\lambda$. In this limit, the condensate is in the harmonic-oscillator 
ground state in the $z$ direction; integrating out the $z$-dependence, 
one obtains the effective coupling constant
$
\tC = C\lambda^{1/2}/\sqrt{2 \pi}. 
$
As mentioned above, we choose to state our results in terms of 
the chemical potential $\mu$, which is a monotone function of $\tC$; 
this will simplify some of the calculations and make some relations 
more clear. 

In the two-dimensional case it is known that the doubly quantized vortex is 
dynamically unstable only in certain intervals of the coupling strength 
\cite{pu1999}. 
The first instability sets in at zero coupling, as can be 
seen by considering the restricted eigenvalue problem, Eq.\ 
(\ref{twobytwomatrix}), for the lowest-lying eigenstates of the 
noninteracting problem. 
Thus we take $\varphi_1=(\phi_{400},0)^T$
and $\varphi_2=(0,\phi_{000})^T$, where $\phi_{l,n_r,n_z}(r,\theta,z)$ 
are the harmonic-oscillator eigenstates in cylindrical coordinates,
\beq
\phi_{l,n_r,n_z}(r,\theta,z) = 
\frac{\lambda^{1/4}}{\sqrt{2^{n_z} n_z! \sqrt{\pi}}}H_{n_z}(z) 
e^{-\lambda z^2/2}
\sqrt{\frac{n_r!}{(n_r+l)!}}L_{n_rl}(r^2)\left(re^{i\theta}\right)^l
e^{-r^2/2},
\enq
where $H_n$ is a Hermite polynomial and
\beq
L_{n\alpha}(x)=\sum_{j=0}^{n}(-1)^j \binom{n+\alpha}{n-j} 
\frac{1}{j!}x^j
\enq
is a generalized Laguerre polynomial.
The two eigenstates $\varphi_1$ and $\varphi_2$ 
are degenerate with Bogoliubov eigenvalue 
$\omega=2$ in the 
noninteracting limit,
but have opposite norms. Thus, for any finite coupling strength $\tC$, the 
Bogoliubov eigenvalues calculated from Eq.\ (\ref{twobytwomatrix}) 
turn complex. 
In order to determine the point where the system turns dynamically 
stable again, one has to consider the contributions from 
states with higher $n_r$ \cite{jackson2005}. 

\begin{figure}[ht]
\includegraphics[width=0.5\columnwidth]{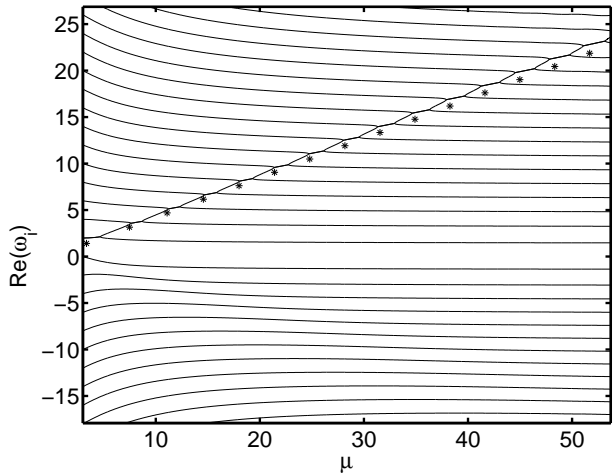}
\caption[]{Energy levels in two dimensions for $m=2$.
 The $*$ represent the analytical approximation to the crossing of the core
 mode with the trap modes shown in Eq.\ (\ref{eqn:analytic_approx}).}
\label{fig:levels2D}
\end{figure}

Figure \ref{fig:levels2D} shows the numerically computed 
real parts of the Bogoliubov 
eigenenergies as functions of the chemical potential $\mu$
in two dimensions. 
According to Eq.\ (\ref{complexmodes}), the merging of two lines into 
one signifies the onset of instability, as 
two real energies become a complex conjugate pair. 
It is seen how the successive instability windows arise from the coupling 
of one mode, whose energy increases monotonously, with successive 
radially excited modes. This mode has no radial nodes and has negative 
norm in the intervals where its energy is purely real; we refer to it as 
the core mode since it is confined to the 
core of the doubly quantized vortex and
causes 
a splitting of the 
vortex into two. Its physical significance is that the doubly quantized 
vortex is {\it energetically} unstable 
towards splitting,
which furthermore causes dynamical instability 
in certain windows of parameter space.
As the coupling increases, the energy of the vortex is negligible 
compared with the
energy contributions from the interaction and trapping potential. 
Therefore all the Bogoliubov energy levels approach 
the excitation energies for a two-dimensional condensate in the 
Thomas-Fermi limit
\cite{stringari1998,Zambelli1998}, except for the core mode
whose
(real part of its) energy increases with increasing 
$\mu$.
In fact, the dependence of the core mode energy on the chemical 
potential is approximately linear. 
This can be understood intuitively as follows.
If the core mode is
populated, the doubly quantized vortex 
is split into two. The energy of the core mode corresponds to the 
frequency of the precession of the two vortices (since it determines 
the rate of change of the relative phase of the condensate wave function 
and the Bogoliubov mode). 
Each of the two vortices moves with the local velocity, which 
is given by the velocity field from the other vortex and is 
proportional to the inverse of their separation \cite{pethick2001}. 
The angular 
frequency of this precession is thus
$\omega_{\rm prec} = \hbar/(m r_{12}^2)$. Inserting for $r_{12}$ the
healing length $\xi$, which is the size of a vortex core, one obtains 
indeed $\hbar\omega_{\rm prec} \propto \mu$ 
\cite{Nilsen06}.

\begin{figure}[htb]
\includegraphics[width=8cm]{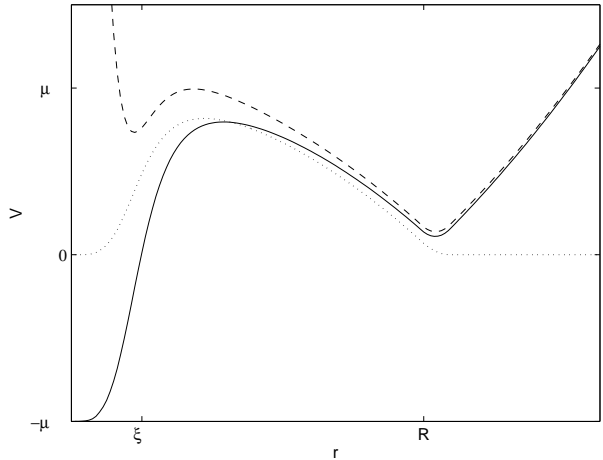}
\caption{Effective potentials entering the Bogoliubov equation for 
a doubly quantized vortex. 
Full line represents the diagonal potential $V_d$, the dashed line
represents the diagonal potential with the addition of a 
centrifugal term, $V_d+4/r^2$, and the dotted line is the 
off-diagonal potential $V_c$. The potentials are 
defined in Eq.\ (\ref{cyl_bog_matrix}).}
\label{illufig}
\end{figure}

A more quantitative estimate of the core mode energy can be obtained 
in the limit of a large condensate. 
Consider the Bogoliubov equation 
(\ref{bogoliuboveqn}) with Eq.\ (\ref{cyl_bog_matrix}) inserted for 
the matrix operator $B$,
and neglect the $z$ depencence.
We do the analysis for a general set of quantum 
numbers $M, m$ before specializing to the currently relevant case 
$M=m=2$.
The diagonal and off-diagonal effective 
potentials $V_d(r)$ and $V_c(r)$ defined in Eq.\ (\ref{cyl_bog_matrix})
are sketched in Fig.\ \ref{illufig}.
Since the norm of the core 
mode $\varphi$ is negative, it is dominated by its lower 
index $v$.
The lowest-lying eigenstates of the diagonal potential $V_d$ are 
concentrated to
the vortex core, i.~e.\ the innermost potential well of the 
potential $V_d$. 
This justifies the term ``core mode'' for this mode, and we therefore 
denote the amplitude by $\vcore$ and the corresponding energy by $\Ecore$. 
The corresponding upper amplitude $u_{\rm core}$ experiences an
additional centrifugal potential which pushes it away 
from the origin. As a result, the overlap between $u_{\rm core}$ 
and $\vcore$ weighted by $V_c$ is exceedingly small and we can neglect 
the off-diagonal term in the Bogoliubov equation for this mode. 
Moreover, in the limit of a large condensate, 
the extent of the function $\vcore(r)$ is of order $\xi$ and the trapping 
potential contributes little to its eigenenergy. Hence the 
Bogoliubov equation for the core mode can 
be well approximated by the diagonal term neglecting the external potential,
\beq
\label{trilbod_gpe}
  \left[\frac{\hbar^2}{2m}\left(-\frac{\partial^2}{\partial r^2}-
  \frac{1}{r}\frac{\partial}{\partial r} +\frac{(M-m)^2}{r^2}
  \right)
  -\mu+2U_0|\Psi(r)|^2\right]\vcore(r) = -\Ecore \vcore(r).
\enq
The condensate wave function $\Psi$ can be written as the product of the 
square root of the density away from the core, $n_0$, and a core 
function $f(r)$ describing how $\Psi$ goes to zero due to the centrifugal 
force: $\Psi(r) = \sqrt{n_0}f(r)$. 
The density $n_0$ is unaffected by the trapping potential at 
distances $r$ comparable to $\xi$, and is assumed constant; it is related 
to the chemical potential by $\mu=U_0n_0$. Rescaling the length, 
$r=\xi x$, results in the equation
\beq
\label{trialbog}
\left[\frac{1}{2}\left(-\frac{\partial^2}{\partial x^2}-
  \frac{1}{x}\frac{\partial}{\partial x}
  +\frac{(M-m)^2}{x^2}
  \right)-\frac12+f(x)^2\right]\vcore(x) = -\frac{\Ecore}{2\mu} \vcore(x).
\enq
Already from here one can see that since all parameters are scaled 
away from the left-hand side, $\Ecore/\mu$ must be a constant, i.~e.\
the core mode energy is proportional to the chemical potential. A 
variational quantitative estimate can be obtained as follows.
Assuming a general condensate with angular momentum $M$ and modeling 
$f(x)$ as a linearly increasing function  
cut off at the position $x=b$, we obtain by variational means 
$b=M\sqrt{6}$ \cite{corefootnote}. With this assumption for the function $f$, 
the solutions $\vcore$ to Eq.\ (\ref{trialbog}) are harmonic-oscillator 
eigenfunctions and the eigenenergies
come out as
\begin{equation}
  \frac{\Ecore}{\mu}=-(2n_r+(M-m)+1)\frac{2}{M\sqrt{3}}+1.
  \label{eqn:analytic_approx}
\end{equation}
We are interested in the core modes with $n_r=0$. 
This energy is indeed positive for 
$M \geq 2$ and $m > 0$, i.~e.\ of opposite sign compared to the norm. 
(Recall that we could equally well have considered the
corresponding 
Bogoliubov mode with $m\to -m$ and the upper component $u$ confined to the 
core; it has positive norm and negative energy.)

The above estimate is readily checked against the numerical result 
for a two-dimensional trapped condensate in the strong-coupling regime. 
We read off the numerically calculated core mode energy at the arbitrarily 
chosen point $\mu=40\hw$.
For the $m=2$ core mode in a condensate with $M=2$, we obtain 
the variational estimate $\Ecore=0.423\mu$, while the numerical 
result is $\Ecore=0.438\mu$.
For $m=3$ and $M=3$, we obtain variationally $\Ecore=0.615\mu$ and numerically 
$\Ecore=0.665\mu$, and for the $m=4$ core mode in a $M=4$ condensate the  
variational result is $\Ecore=0.711\mu$ and the numerical result is 
$\Ecore=0.781\mu$. 

The positions of the successive instability windows can now be 
estimated by calculating the crossing of the core mode 
with quadrupole modes with increasing radial quantum numbers $n_r$. 
The energies of the latter will in the limit of strong coupling not 
depend on the multiply quantized vortex in the center of the 
condensate, so we can use the values calculated for a nonrotating 
two-dimensional condensate
\cite{stringari1998,Zambelli1998}
\begin{equation}
  \omega_{n_r,m}=\hw \sqrt{2 n_r^2+2 n_r m +2 n_r+ m}.
\end{equation}
The variational estimates for the crossings, i.~e.\ the points
$\mu$ where $\Ecore=\omega_{n_r,m}$ for $m=2$ and some $n_r$,
are indicated in Fig.\ \ref{fig:levels2D}.
It is possible to proceed and calculate the overlap between the core 
and quadrupole modes, which yields the widths of the unstable windows 
and the imaginary parts of the eigenfrequencies. However, this is 
impossible in practice, since the result 
will be extremely sensitively dependent on the width of the core mode, 
which is a variational parameter. We will therefore not pursue 
this analysis.

\section{Instability regions for a three-dimensional condensate}
\label{sec:results}

\begin{figure}[ht]
\includegraphics[width=0.45\columnwidth]{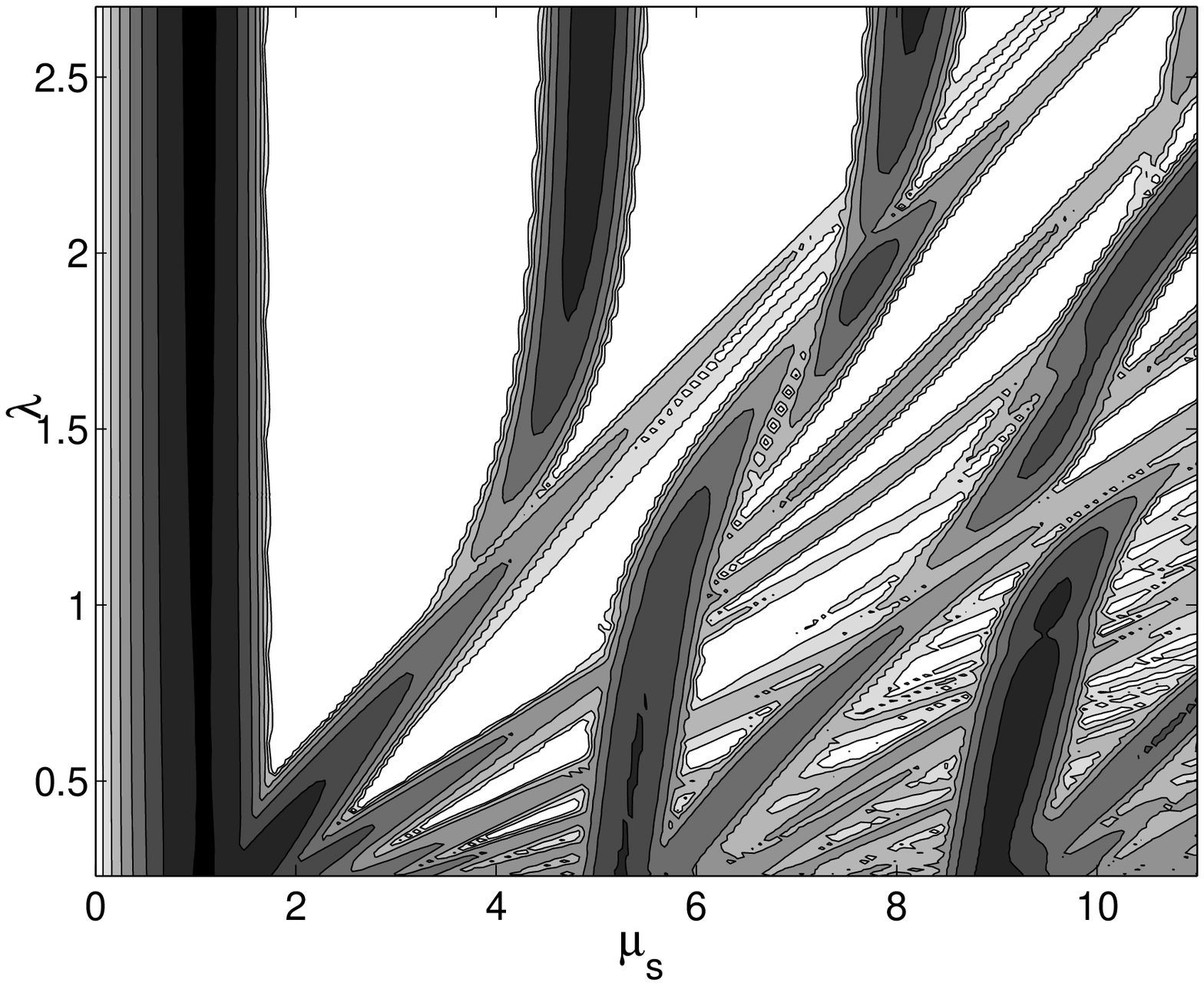}
\includegraphics[width=0.45\columnwidth]{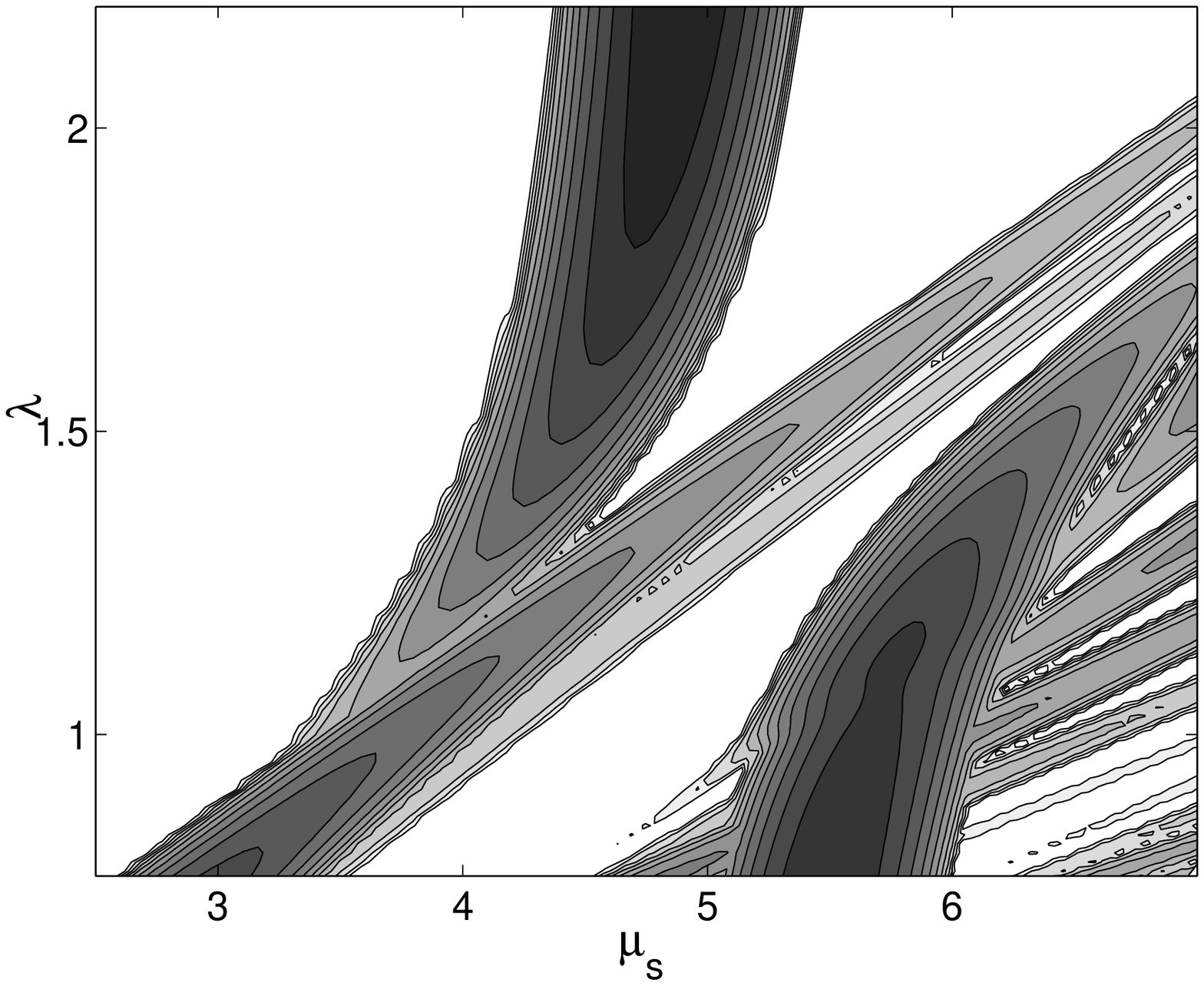}
\caption[]{Unstable regions for a doubly quantized vortex 
in the parameter space of trap anisotropy $\lambda$ and 
shifted chemical potential $\mu_s$.
The shading indicates the 
largest of 
the imaginary parts of the Bogoliubov eigenvalues. White is zero which 
means that the condensate is stable.
\label{fig:phasediag}}
\end{figure}

Figure \ref{fig:phasediag} contains the main result of this study. 
Displayed is the largest imaginary part of the Bogoliubov eigenenergies 
as a function of trap anisotropy $\lambda$ and shifted chemical 
potential $\mu_s$. 
The latter is defined as the chemical potential downshifted by its 
value in the noninteracting limit,
\beq
\mu_s = \mu - \lambda/2 - 3,
\enq
and it will play the role of our coupling parameter in the following.
All the unstable modes are in the 
$m=2$ sector, there are no unstable modes with quantum number $m\neq 2$.
An instability with $m=2$ 
corresponds to a splitting of the doubly quantized vortex line into 
two single vortex lines, as can be seen by considering the density 
profile of a superposition of $m=2$, $m=0$ and $m=4$ radial 
eigenfunctions: In the noninteracting limit the condensate 
wavefunction at some time instant $t$ can be written
\beq
\label{superposition_example}
\Psi(x,y,z,t) = \left(\frac{\lambda}{\pi}\right)^{1/4}\left[
  \frac1{\sqrt{2}}(x+iy)^2+\eta \frac1{\sqrt{4!}}(x+iy)^4 + \eta^*
\right]e^{-(x^2+y^2+\lambda z^2)/2-\mu t},
\enq
where $\eta$ is a complex, time-dependent amplitude. The quantity 
within brackets can for small $x$, $y$ be written as
$[x-x_0+i(y-y_0)][x+x_0+i(y+y_0)]$, which describes two singly 
quantized vortices at opposite sides of the $z$ axis, with 
${\rm Re} \eta = y_0^2-x_0^2$ 
and ${\rm Im}\eta=2x_0y_0$.
Note that when the coordinate along the abscissa 
in Fig.\ \ref{fig:phasediag} is chosen to be the shifted 
chemical potential $\mu_s$, 
the instability regions in the large-$\lambda$ limit appear as 
vertical stripes, as we shortly discuss.
The instability regions bounded by 
straight diagonal lines are
instabilities between negative energy states $n_r=0,n_z=n_n$ and
positive energy states $n_r=0,n_z=n_p$ with $n_n-n_p$ even. 
They correspond to instabilities with axial nodes. 
In the next section we 
shall study the instabilities in greater detail.
Of course, one has to keep in mind that the quantum numbers 
$n_r,n_z$ only make sense in the weak coupling 
limit. Their meaning is lost in the strong-coupling 
limit, where the $r$ and $z$ dependence is no longer separable, 
but it remains a convenient way to label the states.

\begin{figure}
\includegraphics[width=\columnwidth]{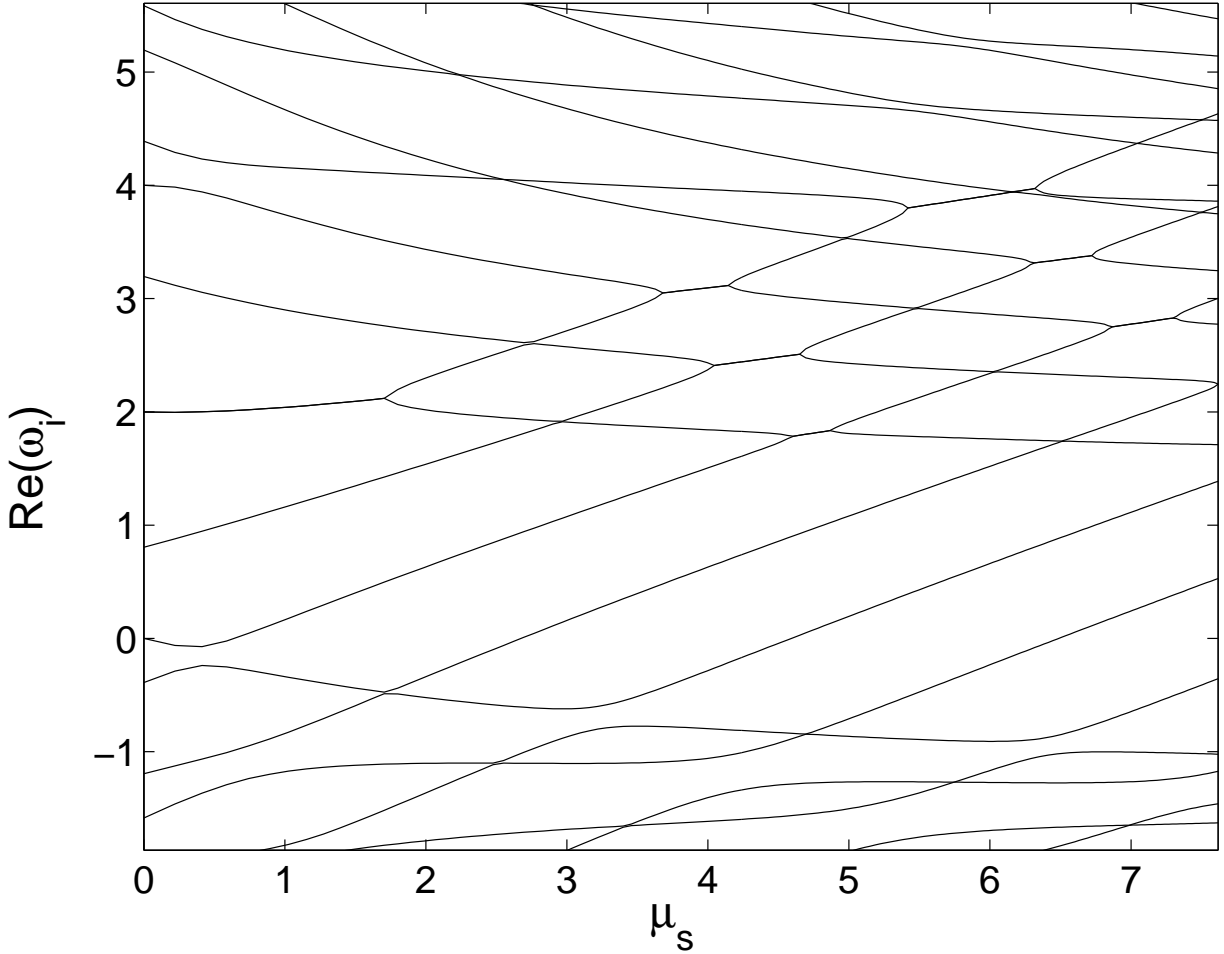}
\caption[]{Real parts of the Bogoliubov eigenenergies as functions 
of the shifted chemical potential $\mu_s$ for the trap 
anisotropy $\lambda=1.2$, i.~e., an almost isotropic trapping 
potential.
\label{fig:levelskule}}
\end{figure}

The alternating stability and instability regions known from previous
two-dimensional studies \cite{pu1999,mottonen2003,jackson2005} 
and discussed in Sec.\ \ref{sec:twodim} are clearly 
seen in the pancake shaped, large $\lambda$ limit as vertical stripes. 
We refer to these as 2D instabilities, since they arise from 
dynamics in the plane. 
When $\lambda$ approaches unity 
from above, one sees how the 2D instability regions become distorted.
In fact, the distortion of the second vertical stripe can be explained 
as an avoided crossing phenomenon. As discussed in Sec.\ \ref{sec:twodim}, 
it is the 
crossing of the $n_r=1$, $n_z=0$ mode with the $n_r=0$, $n_z=0$ core 
mode in the pancake shaped limit that gives rise to the second vertical stripe. 
On the other hand, in the cigar shaped limit the instability of the 
$n_r=0$, $n_z=2$ mode with the $n_r=0$, $n_z=0$ core mode appears as 
a diagonal band in the lower part of Fig.\ \ref{fig:phasediag}, as 
we discuss in Sec.\ \ref{sec:cigar}. The two modes ($n_r=1$,$n_z=0$) 
and ($n_r=0$, $n_z=2$) have the same symmetry and positive norm, and will 
mix when their energy is similar, giving rise to an avoided crossing 
around $\lambda=1$.
This avoided crossing of the real energy levels 
is shown in Fig.\ \ref{fig:levelskule} for $\lambda=1.2$.
The two modes in question start off at $\mu_s=0$ as 
harmonic-oscillator eigenstates with energies $\omega_{n_r=1,n_z=0}=4$ 
and $\omega_{n_r=0,n_z=2}=2+2 \lambda=4.4$, 
respectively. The lines can be
seen to undergo an avoided crossing around $\mu_s=0.5$; the exact 
position of this crossing
depends on $\lambda$. Furthermore, both of these eigenmodes become
unstable when they cross with the core mode, the negative-norm 
eigenmode $n_r=0,n_z=0$. 
The result is
an avoided crossing of the instability regions.
In the 
second panel of Fig. \ref{fig:phasediag} the instability region in
the high $\lambda$ region corresponds to $n_r=1,n_z=0$; this goes
continously over to $n_r=0,n_z=2$ in the lower left corner, while
we find
the $n_r=1,n_z=0$ state again as the strong instability
between $5\lesssim \mu_s\lesssim 6$ in the lower part of the figure.

\section{Instability in the cigar shaped limit}
\label{sec:cigar}

When $\lambda\ll 1$, the condensate obtains an elongated, cigar-like 
shape. In this limit 
the lowest-energy excitations are in the $z$ 
direction, since their energy separation is 
$\lambda\hw$ while the 
radial excitation energy is equal to $\hw$. 
The problem is thus one-dimensional as long as radial excitations 
can be neglected. 
The instability regions carrying different quantum numbers $n_z$ 
are seen to spread 
out like a fan in the lower part of Fig.\ \ref{fig:phasediag}.
The analysis in 
the previous paragraph still holds for the $m=2, n_r=n_z=0$ 
instability, which sets in at zero coupling 
also in the cigar shaped limit. 
Because of parity, this mode does not mix with the modes that have 
odd axial quantum numbers $n_z$ and we now deal with those separately. 
Consider again the Bogoliubov equation restricted to the space spanned 
by two nearby modes, Eq.\ (\ref{twobytwomatrix}).
Assuming that perturbation theory holds so that harmonic-oscillator 
eigenfunctions can be used, we insert into Eq. (\ref{twobytwomatrix}) 
the trial Bogoliubov amplitudes $\varphi_1=(\phi_{401},0)^T$ and 
$\varphi_1=(0,\phi_{001})^T$, where we remind the reader that the 
harmonic-oscillator eigenfunctions $\phi_{l,n_r,n_z}$ are indexed 
with the azimuthal, 
radial, and axial quantum numbers in turn.
The calculation shows that the 
mode is unstable when
\beq
\frac{512\pi^{3/2}}{17\sqrt{2}+8\sqrt{3}} < 
\frac{\tC}{\lambda} <
\frac{512\pi^{3/2}}{17\sqrt{2}-8\sqrt{3}}.
\enq
We rephrase this in terms of the shifted chemical potential, which 
in the weak-coupling limit varies as
$\mu_s=3\tC/[3(2\pi)^{3/2}]$. There results
\beq
\frac{12}{17+4\sqrt{6}} < \frac{\mu_s}{\lambda} 
< \frac{12}{17-4\sqrt{6}}.
\enq
Thus, the fan-like structure in the lower part of 
Fig.\ \ref{fig:phasediag} should be bounded from the left by the ray
$\lambda\approx 0.6002\mu_s$. Although the boundary 
of this instability region appears fairly straight, its slope is 
about half the value predicted by this weak-coupling 
analysis. This is because the instability sets in at a coupling 
strength where many more harmonic-oscillator eigenfunctions are 
already mixed into the relevant Bogoliubov modes.
Going to the limit of small $\lambda$ and small $\mu_s$ does not help, 
since the above analysis indicates that the instability 
sets in only when the interaction energy, which is proportional to 
the shifted chemical potential $\mu_s$, is of the 
order of the the harmonic-oscillator level spacing 
$\lambda\hw$.

\begin{figure}[ht]
\includegraphics[width=0.5\columnwidth]{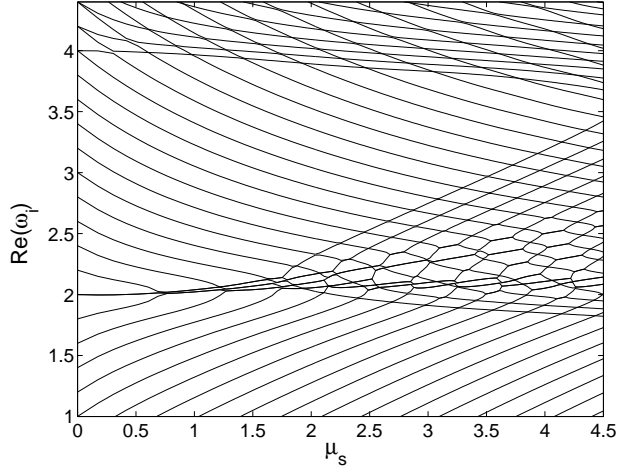}
\caption[]{Real part of the Bogoliubov energy spectrum for the case 
$\lambda=0.2$, i.~e., in the limit of a cigar shaped condensate.}
\label{fig:levelscigar}
\end{figure}

Figure \ref{fig:levelscigar} shows the real parts of the Bogoliubov 
eigenenergies as functions of the coupling for the fixed anisotropy 
$\lambda=0.2$.
It is clearly seen that the radially excited states, in the upper part 
of the figure, are well separated from the lowest axially excited states 
and thus they do not affect the dynamics in the weak-coupling limit.
Again, the simplest way to determine the quantum numbers of a particular 
mode is to inspect its energy in the noninteracting limit. 
The two modes that in the noninteracting limit have energies
$\omega=(2\pm\lambda)\hw=(2\pm 0.2)\hw$ are the ones with quantum number 
$n_z=1$. The initial linear behavior of the energies can be 
obtained from the harmonic-oscillator eigenfunction analysis in the 
previous paragraph, but the curvature of the lines becomes 
important, and as a result the levels merge at a higher chemical 
potential than predicted by the weak-coupling analysis.

As discussed above, the merging of the two lines into 
one signifies the onset of instability, where according to Eq.\ 
(\ref{complexmodes}) the two 
real energies become a complex 
conjugate pair. At a somewhat larger
$\mu_s$
the modes with 
$n_z=2$ 
merge and 
become complex. The levels split apart and become purely 
real again for a larger value of
$\mu_s$. 
As long as the system is approximately one-dimensional in 
the sense that radial excitations are well separated from the axial 
ones, the curves presented in Fig.\ \ref{fig:levelscigar} are 
universal and are only dilated by a factor $\lambda$ as the 
anisotropy is changed. As a result, the instability boundaries appear 
as straight lines that give rise to the fan-shaped structure 
in Fig.\ \ref{fig:phasediag}.

\begin{figure}[ht]
  \includegraphics[width=0.5\columnwidth]{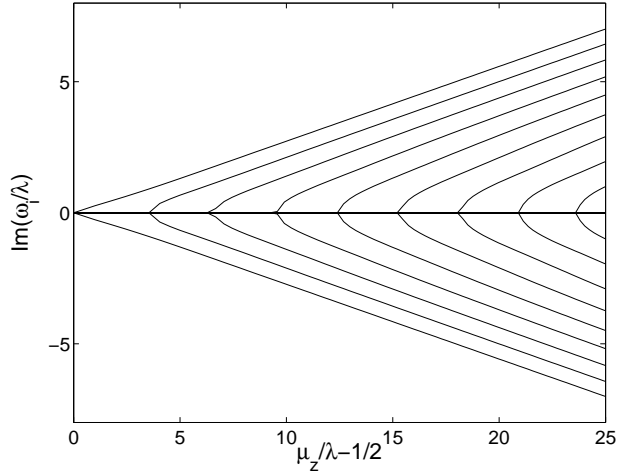}
  \caption[]{Imaginary parts of the Bogoliubov eigenvalues, 
    ${\textrm Im}(\omega_i)$, 
    in the limit of weak interaction and strongly cigar-shaped geometry, 
    calculated using the one-dimensional model defined in Eq.\ 
    (\ref{onedimmodel}).}
  \label{fig:sigar_model}
\end{figure}
In the extremely cigar shaped limit, we can 
completely ignore 
the excitations in the radial direction. 
In this limit the problem simplifies enough 
to allow for analytical calculation of the mode frequencies; 
however, as we shall see, the model is only accurate for very weak 
coupling. 
The Bogoliubov equation will in this case be an
eigenvalue problem of two coupled one-dimensional 
differential equations. 
Inserting the lowest radial harmonic-oscillator eigenfunctions for 
$\Psi$, $u$ and $v$ in the Bogoliubov matrix operator, 
Eq.\ (\ref{cyl_bog_matrix}), integrating out the radial direction and 
redefining units by
putting $\widetilde{z}=(\lambda z)/\sqrt{2\mu_s}$ and
the axial condensate function
$\Psi(z)=\mu_s\widetilde{\Psi}(z)/C$
we obtain 
\begin{eqnarray}
\left[
  \begin{array}{c c}
      -\frac{1}{(\frac{\mu_s}{\lambda})^2}\frac{\partial^2}{\partial
       \tilde{z}^2}+2\frac58|\tilde{\Psi}|^2-1 &
      \frac{1}{\sqrt{6}}|\tilde{\Psi}|^2\\
      -\frac{1}{\sqrt{6}}|\tilde{\Psi}|^2&
      -\left(      -\frac{1}{(\frac{\mu_s}{\lambda})^2}\frac{\partial^2}{\partial
        \tilde{z}^2}+2\frac23|\tilde{\Psi}|^2-1\right)
  \end{array}
  \right]
\left[
  \begin{array}{c}
    u\\
    v
  \end{array}
  \right]
=
\frac{\omega_i}{\mu_s}
\left[
  \begin{array}{c}
    u\\
    v
  \end{array}
  \right].
\label{onedimmodel}
\end{eqnarray}
If we furthermore assume that the
Thomas-Fermi (TF) 
approximation \cite{pethick2001} for the condensate wavefunction 
is valid in the $z$ direction, the peak value of $\tilde\Psi$ is 
just equal to 1, and
the only remaining 
parameter in the equation is $\mu_s/\lambda$. Thus the points at
which the different axial modes become complex are given by
$\mu_s/\lambda={\rm const}$, i.e, straight lines in figure
\ref{fig:sigar_model}. Furthermore, in the limit $\mu_s/\lambda\rightarrow
\infty$ we can neglect the derivative of the function at the center of
the trap and the eigenvalues
$\omega_i/\mu_s$ have to 
approach
the eigenvalues of the matrix
\begin{equation}
 A=\left[
    \begin{array}{c c}
      \frac{1}{4} & \frac{1}{\sqrt{6}}\\
      -\frac{1}{\sqrt{6}} &-\frac{1}{3}
    \end{array}
    \right].
\end{equation}
 The result is $\omega_i = 
 \mu_s(-1/24\pm i\sqrt{47}/24)- O(\lambda)$. 
In the actual 
three-dimensional situation we have seen that the imaginary 
parts of the frequencies rise to a maximum value and then decrease to 
zero; this behavior is not predicted by the 1D model and is thus a 
three-dimensional effect. In order to estimate the maximum values of the 
imaginary parts of the frequencies, we observe that 
the radial dynamics is expected to begin to matter when $\mu_s$ exceeds unity, 
so putting $\mu_s=1$ gives an estimate
${\rm Im}(\omega_i)\sim 0.3 $. 
The actual value from the three-dimensional calculation turns
out to be 0.14 for the maximum of the first imaginary frequency, with a 
slight dependence on  $\lambda$. 
We conclude from the one-dimensional calculation that the periodic regions 
of stability and the maximum of the imaginary part of the eigenvalue is an 
effect arising from the dynamics in the radial direction.

It is again instructive to visualize the instability by considering the 
shape of the wave function $\psi(r,\theta,z)$ with a small 
admixture of the Bogoliubov amplitudes $u$ and $v$, as we did in 
Eq.\ (\ref{superposition_example}). One sees that 
the $m=2, n_z=0$ modes simply correspond to a straight splitting 
of the doubly quantized vortex into two. The $m=2, n_z=1$ mode 
corresponds to two vortices that split at the edges of the 
condensate, at large $|z|$, but are joined at $z=0$, thus forming 
an X-shaped structure.
In experiments one will generally be in a regime where more than one
mode is unstable. This will result in a intertwining of two
vortices.
This was studied numerically 
in
Refs.\ \cite{mottonen2003,mottonen2006,mateo2006}.
The splitting was found to nucleate in certain intervals of $z$, 
corresponding in the present picture to a high quantum number $n_z$, as 
is expected for strong coupling. 
It was also proposed in 
Refs.\ \cite{mottonen2003,mateo2006} that the criterion for local splitting 
can be found from a local-density approximation of sorts, by 
treating the elongated condensate as a stack of two-dimensional 
slices. If the local density integrated over the 
$x$-$y$ 
plane matches 
the instability criterion for the two-dimensional system, an 
instability 
can be nucleated at that point. This kind of analysis 
presumably holds in the limit of a large condensate, where 
local-density approximations are expected to hold.

\section{Experimental lifetime of a doubly quantized vortex}
\label{sec:experiment}

In the experiment carried out by Shin {\it et al.} \cite{shin2004}, 
a doubly quantized vortex was topologically imprinted in a 
$^{23}$Na condensate and its subsequent decay was tracked by 
observing time-of-flight density profiles. It was 
argued in Ref.\ \cite{mottonen2006} that the initial 
occupation of the dynamically unstable modes, which as we have seen 
have quadrupole symmetry, is mainly because of gravitational sag 
in the trap during topological imprinting, which produces a 
quadrupolar deformation. In order to compute the lifetime of a 
doubly quantized vortex as observed in the experiments, 
one has in principle to 
model the full dynamical process including the initial seeding of 
the unstable modes, their growth and mixing, 
nonlinear effects that occur once the mode occupation becomes 
appreciable, migration of the density fluctuations along 
the vortex axis \cite{mateo2006}, and finally the expansion 
of the atom cloud before observation of two separate density 
depressions. However, it is clear that the main contribution 
to the lifetime is given by the rate of exponential growth of 
the unstable modes. This rate is just the maximum imaginary part 
of the complex eigenvalues (MCE) of the condensate at that 
particular point in phase space.
The lifetime depends only logarithmically on the initial 
mode population and is thus insensitive to the seeding 
process.
Whether nonlinear processes during the latter stages of 
the decay can appreciably affect the dependence on coupling strength 
is more of an open question;
this remains to be investigated and 
we shall see in this section that the overall parameter dependence 
seems to be very well described by the sole parameter that is the 
MCE.

\begin{figure}[ht]
  \includegraphics[width=0.5\columnwidth]{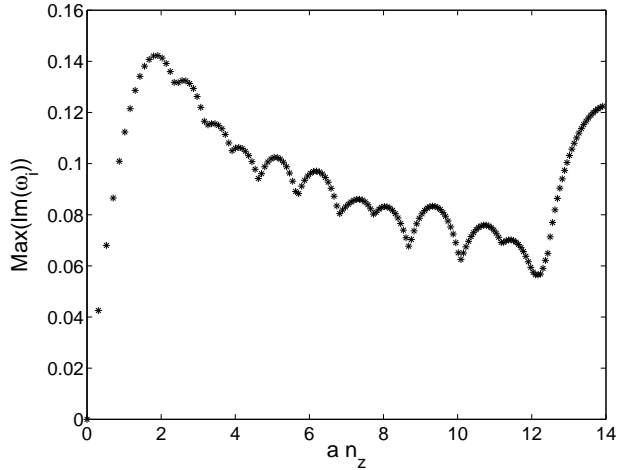}
  \caption[]{The maximal imaginary part of the complex eigenvalues 
    for a condensate with 
    a doubly quantized vortex in the cigar shaped limit with 
    trap anisotropy $\lambda=0.2$, as a function of two-dimensional 
    coupling strength. The two large peaks are termed 2D instabilities 
    since they do not depend on the coordinate along the vortex line, 
    while the unstable modes connected with the smaller peaks have 
    nodes in the axial direction.}
  \label{fig:max_omega_exp}
\end{figure}

The experiment by Shin {\it et al.} \cite{shin2004} was done with aspect
ratios $\lambda$ ranging 
from $1/100$ to $1/20$. This is clearly in the cigar-shaped domain. In
this region the unstable modes are quadrupole modes with 
different numbers of axial
nodes, as we have seen in the preceding section. 
Figure \ref{fig:max_omega_exp} shows the value of the 
MCE for  $\lambda=0.2$, as a function of the 
effective two-dimensional coupling strength $an_z$, which is the 
parameter that was used in Ref.\ \cite{shin2004}.
It is defined as 
$an_z=a\int\Psi(r,0)2\pi r dr$, and has to be computed numerically for 
each data point.
Because of numerical limitations, we have used a larger value of 
$\lambda$ than in the experiment, but as we have seen, the most 
important features, such as the position of the 2D instabilities, 
do not change when $\lambda$ is decreased.
In the region between the two 2D instabilities, the most unstable 
mode acquires successively higher numbers of axial nodes. 
The instability gets weaker with
higher density in the region $0 < an_z < 12$, which is precisely the 
parameter interval 
that was scanned in the experiment. 
Clearly, because of a coincidence the experiment was performed in the 
parameter regime lying 
exactly between the two first 2D instabilities. 

As we saw in Sec.\ \ref{sec:cigar}, finding the MCE 
is a purely computational task, since 
a good analytical approximation seems to be difficult to construct. 
Qualitatively we can understand that the MCE decreases
with increasing coupling because the coupling matrix
element is smaller when the
unstable modes have axial nodes.
\begin{figure}[ht]
  \includegraphics[width=0.5\columnwidth]{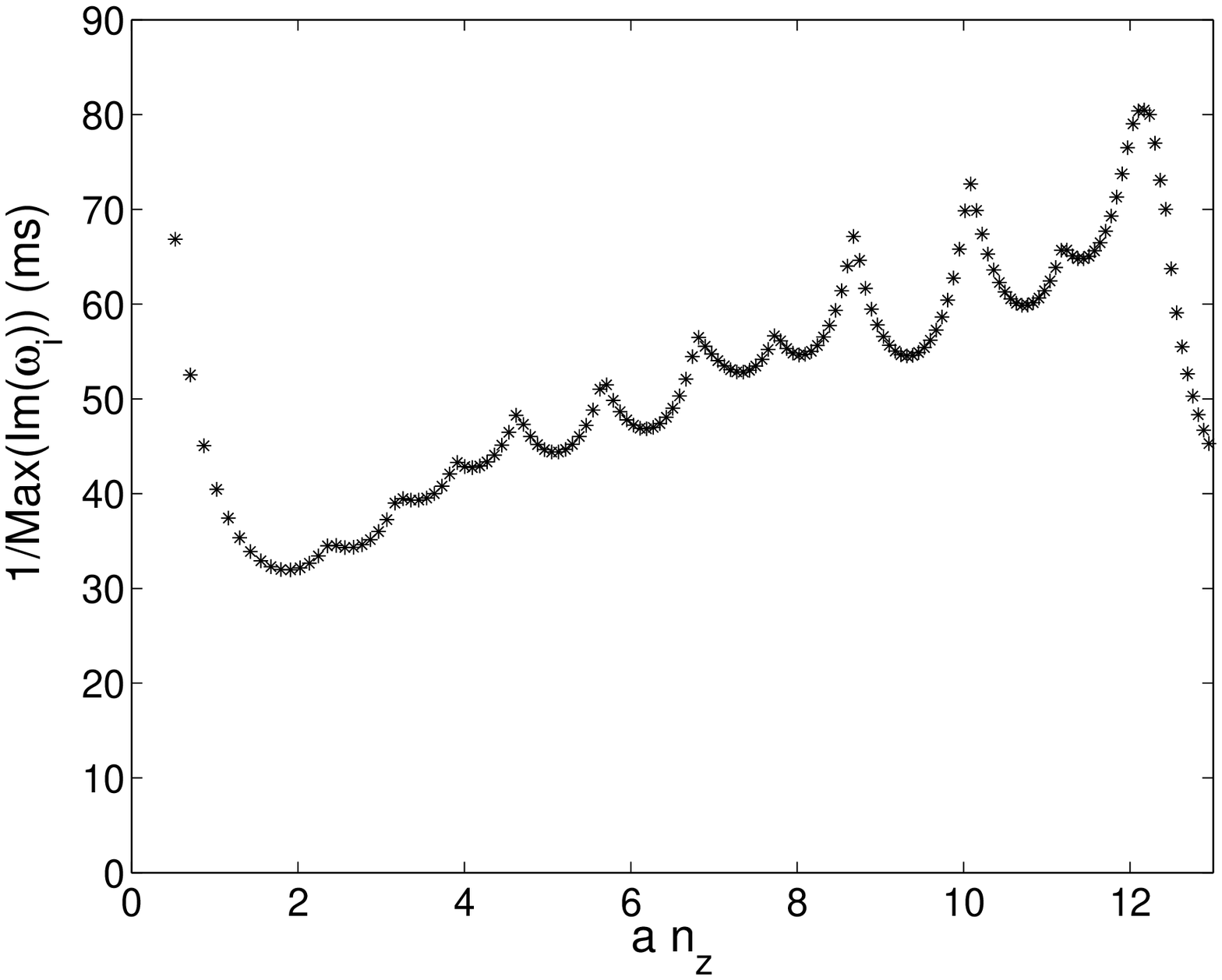}
  \caption[]{Time scale for splitting of a doubly quantized vortex 
    in a cigar shaped condensate with anisotropy $\lambda=0.2$, 
    defined as the reciprocal of the 
    largest imaginary part of the Bogoliubov eigenvalues. Units on 
    the axes are chosen to correspond with the experiment of Ref.\ 
    \cite{shin2004}.}
  \label{fig:max_omega_time_exp}
\end{figure}
To compare with the experiment \cite{shin2004}, we 
identify the lifetime 
of a doubly quantized vortex with the reciprocal of the MCE.
The result is presented in Fig.\ \ref{fig:max_omega_time_exp} 
in the units used in the experiment. 
The numerical result has the correct qualitative behavior, and the 
quantitative scale is also similar to the experimental result.
We conclude that while nonlinear effects and mode mixing may be 
necessary to quantitatively fine-tune the splitting times, it seems 
that simply taking the reciprocal of the MCE is sufficient to 
obtain the parameter dependence of the vortex lifetime both 
qualitatively and roughly quantitatively.

\section{Stability in an anharmonic trap}
\label{sec:anharmonic}
\begin{figure}[ht]
\includegraphics[width=0.5\columnwidth]{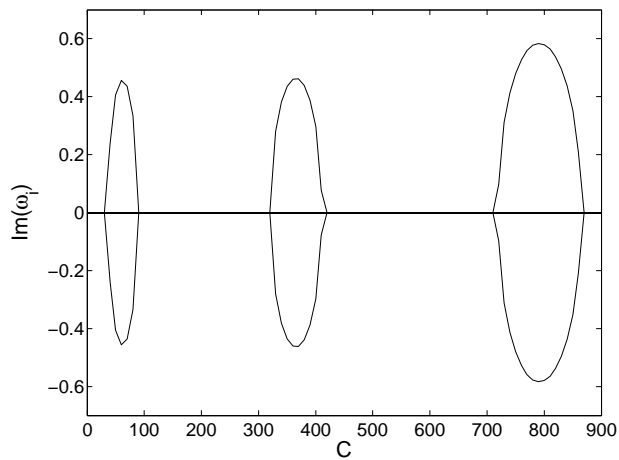}
\caption[]{Imaginary part of the energy spectrum for the two-dimensional case 
in an anharmonic potential with $\alpha=0.92$.}
\label{fig:im_an}
\end{figure}

In the previous sections we have studied a doubly quantized
vortex in a harmonic trap. In this case the vortex becomes unstable
already for 
any finite
value of the coupling $C$. 
We have seen that this is due to the degeneracy of the eigenvalues in the 
harmonic trap. To understand the connection between the energy spectrum
and the instability further, we now present the corresponding 
calculations for an anharmonic
trap. We use the potential
\begin{equation}
  V(r)=\frac{1}{2}r^2+\left(\alpha r\right)^4.
\end{equation}
The regions of instability for this potential are shown in Fig.\
\ref{fig:im_an}. The energies in the noninteracting limit are no longer 
degenerate as they were in the harmonic case: 
For a given anisotropy, the energy of the negative
norm state is below the energy of the positive norm state for weak coupling. 
The nonlinear coupling therefore needs to attain a finite value in order 
for the two real eigenvalues to meet, so that  
the eigenvalues turn complex and the vortex becomes unstable. 
After this point the
situation will be similar to the harmonic case, except that the 
mode frequencies are shifted, which causes a corresponding shift in the 
positions of the unstable regions.

In the weak-coupling region, before the first complex eigenvalue appears, it is
possible to find a rotating frame with rotation frequency $\Omega$
such that all the positive-norm and negative-norm states have positive and
negative energy respectively 
(i.~e., all modes have a positive Krein signature
\cite{Arnold68}). 
If this is the case for all sectors with 
different angular momentum quantum number $m$ 
(which, in fact, it is for certain $\Omega$ and $C$ values), then 
according to Hamiltonian stability theory
\cite{Arnold68,Kapitula04}
the 
doubly
quantized vortex is energetically stable in that rotating frame.
This
certainly implies that the vortex is stable
(both linearly and nonlinearly), both to perturbations of the vortex
state and also to perturbations of the Hamiltonian that are
rotationally symmetric. 
\begin{figure}[ht]
\includegraphics[width=0.5\columnwidth]{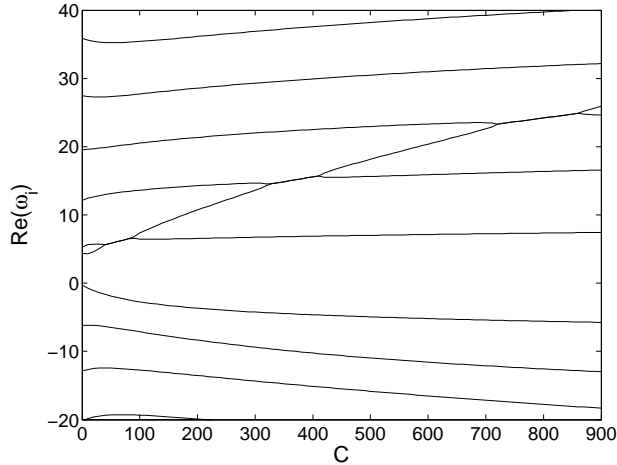}
\caption[]{Real part of the energy spectrum for the two-dimensional case 
in an anharmonic potential with anharmonicity parameter $\alpha=0.92$.}
\label{fig:real_an}
\end{figure}
\begin{figure}[ht]
\includegraphics[width=0.5\columnwidth]{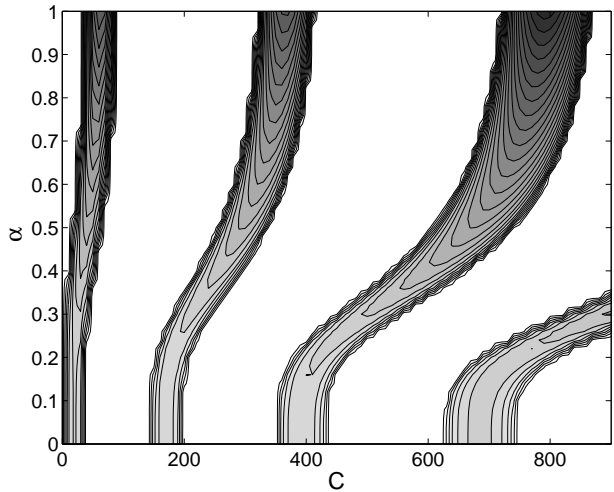}
\caption[]{Dynamical stability regions for a doubly quantized vortex in 
a 
two-dimensional anharmonic trap, as a function of coupling strength $C$ and 
anharmonicity $\alpha$.}
\label{fig:im_max_an}
\end{figure}

\section{Conclusions}
\label{sec:conclusions}
We have in detail studied the excitation spectrum for a doubly quantized 
vortex in a trapped condensate. The
instability regions were studied numerically for a wide range of trap 
shapes and interaction
strengths. For the previously studied 
two-dimensional case, which is expected to described pancake
shaped condensates \cite{pu1999},
we explained the instabilities in terms of level crossings 
between the core mode and the quadrupole modes of the condensates 
and found an analytical approximation for the position of the instability
regions.
A corresponding study of the anharmonic trap was carried out 
in order to point out the
connections and differences between spectral and energetic stability. 
It was found that the doubly quantized vortex is in this case stable 
in the weak-coupling limit, but for stronger coupling the spectrum 
is similar to that for a harmonically trapped condensate.

We have systematically mapped out the regions of instability in a 
three-dimensional trap for a wide range of aspect ratio.
In the cigar shaped regime, as the interaction strength becomes larger 
the unstable modes acquire successively more nodes in the axial 
direction. 
Comparison of the imaginary parts of the computed mode frequencies 
with the results of the experiment performed by Shin
{\it et al.} \cite{shin2004}, was seen to yield qualitative agreement.


\begin{thebibliography}{99}
\bibitem{corefootnote} It may be argued that the condensate wavefunction 
  in the vortex core is better modelled as $r^M$, since this is the true 
  behavior of the exact solution in the limit of small $r$. However, such 
  a trial function is not at all accurate when the wavefunction 
  approaches unity, and we therefore choose to employ the simplest possible 
  trial function, which is a linear one. Indeed, calculations using the 
  ansatz $r^M$ yield results of about the same degree of accuracy as the 
  linear one and we do not report them here.
\end{thebibliography}
\end{document}